# Evaluating the Crowd with Confidence

Manas Joglekar
Stanford University
353 Serra Mall
CA, 94305
manasrj@stanford.edu

Hector Garcia-Molina
Stanford University
353 Serra Mall
CA, 94305
hector@cs.stanford.edu

Aditya Parameswaran
Stanford University
353 Serra Mall
CA, 94305
adityagp@cs.stanford.edu

## ABSTRACT

Worker quality control is a crucial aspect of crowdsourcing systems; typically occupying a large fraction of the time and money invested on crowdsourcing. In this work, we devise techniques to generate confidence intervals for worker error rate estimates, thereby enabling a better evaluation of worker quality. We show that our techniques generate correct confidence intervals on a range of real-world datasets, and demonstrate wide applicability by using them to evict poorly performing workers, and provide confidence intervals on the accuracy of the answers.

## Categories and Subject Descriptors

H.1.0 [**Information Systems Applications**]: Models and Principles—*General*

## 1. INTRODUCTION

A *crowdsourcing system* employs human workers to perform *tasks*, including data processing tasks such as classification and clustering. A major issue in any crowdsourcing system is worker quality: workers can naturally perform some tasks incorrectly, but there are often workers that incorrectly perform more than their share. Some of the low quality workers may not have the necessary abilities for the tasks, some may not have adequate training, and some may simply be "spammers" that want to make money without attempting tasks diligently. Anecdotal evidence indicates that the spammer category is especially problematic, since these workers not only do poor work, but they do a lot of it as they try to maximize their income.

To correct or compensate for poor worker quality, a crowdsourcing system implements some type of *worker quality control* WQC. Typically workers have known identities, so that WQC can identify the poor workers and then possibly take action against them or against their results.

In this paper we focus on one of the most important aspects of WQC: estimating the quality (specifically, error probability) of workers based on their past work. The estimates can then be used either to take action against bad workers (e.g., preventing them from doing future work, paying them less) or for adjusting results (e.g., "downgrading" results from bad workers).

Although there is substantial work on WQC in crowdsourcing systems (see related work section), as far as we know we are the first to estimate worker quality with *confidence intervals* for these estimates. To illustrate the importance of confidence intervals, consider two scenarios. In the first scenario, a particular worker $W_1$ has performed 3 tasks, one of them incorrectly. (Assume that in this case the correct answers are known in advance, i.e., the worker is performing tasks for evaluation purposes only.) Hence, we can estimate the probability that $W_1$ makes a mistake is $1/3$. In Scenario 2, worker $W_2$ has performed 30 tasks, 10 of them incorrectly. In this case we get the same error estimate for $W_2$ as for $W_1$, i.e., $10/30 = 1/3$. However, in the second scenario we are much more confident that worker $W_2$ is making quite a few mistakes. In Scenario 1, perhaps $W_1$ was unlucky or was just distracted, and his/her one incorrect answer is not representative. If we are going to fire workers based on these estimates, it is important that we have "sufficient confidence" in our estimates before we take actions. Thus, we may demand that our $1/3$ error estimate has, say, a 90% confidence interval of size 0.05. Given a requirement like this, we will show how many tasks we must require before making a decision.

Instead of comparing worker results against known correct answers, we focus on schemes that do not require known answers. These schemes are easier to set up, and require no supervision. These well known schemes [7, 8] rely on the frequency of disagreement among workers to estimate error probabilities. In this paper we present a novel way to estimate errors based on disagreements, and we show that our method is as accurate as the previous methods and in addition yields the confidence intervals associated with its estimates.

Determining confidence intervals for a single estimate is a well-studied problem in statistics, with well-known solutions [32]. However, in our scenario, the standard solutions do not apply because we are simultaneously generating confidence intervals for multiple estimates (in this case, the worker error probabilities for each of the workers working on the same task) that are dependent on each other in complex ways.

In this paper, we present a technique that is guaranteed to provide accurate confidence intervals for worker error estimates under some simplifying assumptions (e.g., fixed difficulty tasks, fixed worker error rate during one evaluation period). Even with these assumptions, we will show that our technique provides accurate confidence intervals in practice over real-world datasets that cover a range of crowdsourcing applications, such as peer evaluation, image comparisons, and predicate evaluation. In Section 7, we describe generalizations to our method for other scenarios not captured by our basic technique, such as multiple task types, varying difficulty and categorization.

In summary, we make the following contributions in this paper:

- We provide a new disagreement-based technique for estimating worker quality along with confidence intervals, for three (Section 3) or more workers (Section 4).
- We show that our techniques provide estimates that are as good as Expectation-Maximization [7, 8] and better than Majority based approaches, while in addition providing confidence intervals for those estimates (Section 5).
- We present experimental results based on three different real data sets, and evaluate the accuracy of our estimates and confidence intervals (Sections 3.1, 4.1).
- We show the usefulness of our confidence estimates by applying the techniques to:
  - Estimate accuracy of answers (Section 6.1)
  - Choose between evaluations with multiple workers and multiple tasks (Section 6.2)
  - Choose between workers (Section 6.3)
- Finally, we discuss how to generalize our technique (Section 7).

## 2. MODEL

The crowdsourcing system operates in *phases*. During one phase, a set of workers $W$ perform a set of tasks $T$. Each worker in $W$ performs all $n$ tasks in $T$.

In our base case, we consider tasks that have two possible results, which we will call "Yes" (Y) and "No" (N). The correct outcome is unknown to the evaluation system and the workers. For now we assume we have no a-priori knowledge of the fraction of tasks that have Y as the correct output. (We consider this case in Section 7.)

We assume that a given worker $i$ has an error probability $p_i$. That is, with probability $p_i$ a worker will produce N for a task whose correct result is Y, or will give Y for an N task. Our goal at the end of a phase is to provide an estimate $\hat{p}_i$ for the $p_i$ value of each worker. Our estimate $\hat{p}_i$ will have a confidence interval of half-size $\epsilon_i$ and confidence level $c_i$. The interpretation is that if we consider say 100 phases with similar workers and tasks, in $c_i$ percent of them the true error rate $p_i$ will be between $\hat{p}_i - \epsilon_i$ and $\hat{p}_i + \epsilon_i$.[1] When it is clear what worker we are referring to, we will omit the $i$ subscript and simply use $p$, $c$ and $\epsilon$.

We know that in practice some of these assumptions may not hold. For example, workers may collude so their error rates may not be independent. Worker errors may depend on task "difficulty" or other environmental factors. As mentioned earlier, our assumptions can be relaxed, see Section 7. But it is important to study the base-case solution (with strong assumptions) as it forms the basis of the generalizations. And as also mentioned earlier, the base-case solution can still give quite accurate results in many cases (Section 3.1, 4.1), even if we do not know whether the assumptions hold.

At the end of a phase, we compute the error estimates for our $W$ workers and we take appropriate action, e.g., replacing some of the workers for the next phase. The focus of this paper is on the worker evaluations within a single phase. However, in Section 6.3 we discuss how our estimates can be used to make decisions at the end of a phase, and we show the impact of the decisions across multiple phases.

## 3. 3-DIFFERENCES SCHEME

In this scenario we have three workers with unknown error rates $p_1$, $p_2$ and $p_3$. We are given that these error rates are less than $1/2$. (An error rate larger than $1/2$ is unlikely as it implies that a worker is making worse than random choices.) All three workers do a sequence of $n$ tasks, for which we do not know the correct result, and our goal is to estimate the worker error rates based on the differences in their responses. Intuitively, when a worker disagrees with the majority, it is a sign that the worker may have made a mistake. (In the following section we consider the case of more than 3 workers.)

For each $i$ from 1 to $n$, we define three random variables, $X_{i,12}$, $X_{i,23}$ and $X_{i,13}$, that track the differences among workers in the $i^{th}$ task. For instance, variable $X_{i,12}$ is 1 when workers 1 and 2 agree on the $i^{th}$ task, and 0 otherwise. Let the means of these random variables be auxiliary variables $q_{12}$, $q_{23}$ and $q_{13}$. Since $X_{i,12}$ is 1 when workers 1 and 2 are both right or both wrong, we have $q_{12} = p_1 p_2 + (1 - p_1)(1 - p_2)$. We can get similar equations for $q_{13}$ and $q_{23}$. So we have:

$$q_{12} = p_1 p_2 + (1 - p_1)(1 - p_2) \quad (1)$$
$$q_{23} = p_2 p_3 + (1 - p_2)(1 - p_3) \quad (2)$$
$$q_{13} = p_1 p_3 + (1 - p_1)(1 - p_3) \quad (3)$$

Note that since $p_1, p_2, p_3 < \frac{1}{2}$, we have $p_1 < (1 - p_1)$, $p_2 < (1 - p_2)$ and $p_3 < (1 - p_3)$. Hence by the rearrangement inequality, $q_{12}, q_{13}, q_{23}$ are all greater than $1/2$ and are decreasing in $p_1$, $p_2$, and $p_3$.

Let $X_{12} = \frac{\sum_{i=1}^{n} X_{i,12}}{n}$. Let $X_{23}$, $X_{13}$ be defined similarly. Then each of $X_{12}, X_{23}, X_{13}$ represents an outcome of a Binomial Experiment, which is a sum of $n$ statistically independent Bernoulli Trials. Suppose a Binomial Experiment consists of $n$ Bernoulli trials with mean $p$ each, and the outcome of the experiment is $a$, then it is known [32] that $p$ can be estimated as $\hat{p} = \frac{a}{n}$. And for any confidence level $c$, we can find the $c$-confidence interval for $p$ using the so-called Wilson Score Interval [34].

$$\text{BinInterval}(\hat{p}, c, n) = \frac{\hat{p} + \frac{1}{2n} z_t^2 \pm z_t \sqrt{\frac{\hat{p}(1-\hat{p})}{n} + \frac{z_t^2}{4n^2}}}{1 + \frac{1}{n} z_t^2} \quad (4)$$

where $t = \frac{1-c}{2}$, and $z_t$ gives the $t^{th}$ percentile of the standard normal distribution. The values of $z_t$ are readily available in table form [4, 32].

The half-size of the interval is:

$$\text{Bin}\epsilon(\hat{p}, c, n) = \epsilon \text{ (Interval Half-Size)} = \frac{z_t \sqrt{\frac{\hat{p}(1-\hat{p})}{n} + \frac{z_t^2}{4n^2}}}{1 + \frac{1}{n} z_t^2} \quad (5)$$

For large $n$, the Binomial distribution starts to resemble the Normal distribution, and the terms $\frac{1}{2n} z_t^2$, $\frac{z_t^2}{4n^2}$ and $\frac{1}{n} z_t^2$ become negligible compared to the terms they are being added to. In this case, the end points of our $c$ confidence interval are:

$$\text{BinInterval}'(\hat{p}, c, n) = \hat{p} \pm z_t \sqrt{\frac{\hat{p}(1-\hat{p})}{n}} \quad (6)$$

The half-size of the interval is:

$$\text{Bin}\epsilon'(\hat{p}, c, n) = z_t \sqrt{\frac{\hat{p}(1-\hat{p})}{n}} \quad (7)$$

So if $b_{12}$ is the number of times workers 1 and 2 disagree (out of $n$ tasks), then we can estimate $\hat{q_{12}} = b_{12}/n$. Using Equation 5 (or 7), we can estimate the half-size of the $c$ confidence interval for $\hat{q_{12}}$ as $\epsilon_{12} = \text{Bin}\epsilon(\hat{q}, c, n)$ (or $\epsilon_q \approx \text{Bin}\epsilon'(\hat{q}, c, n)$). The estimates $\hat{q_{23}}$ and $\hat{q_{13}}$, and the half-sizes of their intervals $\epsilon_{23}$ and $\epsilon_{13}$ can be found similarly.

---

[1] With a small number of tasks, $\hat{p}_i$ may not be exactly at the center of the interval.

The $q$ estimates and our desired $p$ estimates are related via Equations 1, 2 and 3, where all variables have hats. We can solve the three equations for $\hat{p_1}$, $\hat{p_2}$ and $\hat{p_3}$:

**Lemma 1.** The value for $\hat{p_1}$ is

$$\hat{p_1} = \frac{1}{2} - \sqrt{\frac{(\hat{q_{12}} - \frac{1}{2})(\hat{q_{13}} - \frac{1}{2})}{2(\hat{q_{23}} - \frac{1}{2})}} \quad (8)$$

The values for $\hat{p_2}$ and $\hat{p_3}$ are computed analogously. □

The proofs for this and the other lemmas in this paper can be found in the extended technical report [16].

Our next task is to compute confidence intervals for $\hat{p_1}$, $\hat{p_2}$ and $\hat{p_3}$. This is more complex than getting confidence intervals for the $q$s because each confidence interval depends on the intervals of the three inter-related variables $X_{12}$, $X_{23}$ and $X_{13}$.

To illustrate our solution, first consider a simpler case. Say random variable $Z = f(X, Y)$ for some function $f$ and random variables $X$ and $Y$. Say $f$ is monotonically increasing in both $X$ and $Y$. Furthermore, say $\hat{x}$ is our estimate for the mean of $X$, with $c$ confidence interval with half-size $\epsilon_x$. Similarly for $Y$ we have $\hat{y}$ and $\epsilon_y$ (same $c$). Now consider the interval $I$ from

$$\min = f(\hat{x} - \epsilon_x, \hat{y} - \epsilon_y) \quad \text{to} \quad (9)$$
$$\max = f(\hat{x} + \epsilon_x, \hat{y} + \epsilon_y). \quad (10)$$

If $X$ and $Y$ were independent variables, then we could say that with probability $c^2$ the true mean of $X$ would be in its interval *and* the true mean of $Y$ would be in its interval *and thus* the true mean of $Z$ is within $I$. (Remember that $f$ is increasing.) Hence, we can use $I$ as the confidence interval for $Z$, with confidence $c^2$ and half-size $(\max - \min)/2$.

Unfortunately, $X$ and $Y$ may not be independent. Hence, we can only provide a bound for the confidence of $I$. In particular, with probability $1 - c$ the mean of $X$ is not in its interval, and with probability $1 - c$ the same is true for $Y$. The probability of the union of these two events is at most $(1 - c) + (1 - c)$. Hence the converse occurs with at least probability $1 - ((1 - c) + (1 - c)) = (2c - 1)$. That is, the mean of $Z$ will be in $I$ with probability $(2c - 1)$ or larger. (We assume that $2c$ is larger than 1.)

If $c$ is close to 1, then $I$ will still have a confidence close to 1. However, if $c$ has a relatively small value, then our bound is less useful. For instance, if $c = 0.80$, our confidence in $I$ is only 0.6.

In Lemma 2 below we show how to compute a confidence bound for our $p$ estimates for the differences scheme (3 workers). The derivation is similar to what we have illustrated, except that there are three variables involved, and composite random variable $Z$ is increasing on two of its parameters and decreasing on the third. In Lemma 3 we then show how this bound can be significantly improved by making two reasonable assumptions.

**Lemma 2.** Define function $f(a, b, c)$ for parameters $a$, $b$ and $c$ to be

$$f(a, b, c) = \frac{1}{2} - \sqrt{\frac{(a - \frac{1}{2})(b - \frac{1}{2})}{2(c - \frac{1}{2})}}$$

Note that $\hat{p_1} = f(\hat{q_{12}}, \hat{q_{13}}, \hat{q_{23}})$. Assume that confidence intervals for $\hat{q_{12}}$, $\hat{q_{23}}$ and $\hat{q_{13}}$ all have confidence $c$ and $c \geq \frac{2}{3}$. The confidence interval for $\hat{p_1}$ ranges from

$$\max_1 = f(q_{12} + \epsilon_{12}, q_{13} + \epsilon_{13}, q_{23} - \epsilon_{23}) \quad \text{and}$$
$$\min_1 = f(q_{12} - \epsilon_{12}, q_{13} - \epsilon_{13}, q_{23} + \epsilon_{23}),$$

where $\epsilon_{12}$, $\epsilon_{23}$ and $\epsilon_{13}$ are the half-sizes of the confidence intervals for the $q$ variables (computed using the large $n$ approximation). The half-size of the $\hat{p_1}$ interval, $\epsilon_1$ is $(\max_1 - \min_1)/2$. The confidence associated with this interval is at least $3c - 2$. The intervals for $\hat{p_2}$ $\hat{p_3}$ are computed in an analogous fashion. □

**Lemma 3.** Consider our three random variables $X_{12}$, $X_{23}$ and $X_{13}$ with mean estimates $\hat{q_{12}}$, $\hat{q_{23}}$ and $\hat{q_{13}}$, and $c$ confidence interval half-sizes of $\epsilon_{12}$, $\epsilon_{23}$ $\epsilon_{13}$ respectively. Also consider a general function $f(a, b, c)$. Let us assume that

- $X_{12}$, $X_{23}$ and $X_{13}$ are normally distributed (large $n$ assumption);
- Function $f$ is locally linear near $a = \hat{q_{12}}$, $b = \hat{q_{23}}$ and $c = \hat{q_{13}}$. That is,

$$f(a, b, c) \approx (a - \hat{q_{12}})d_a + (b - \hat{q_{23}})d_b + (c - \hat{q_{13}})d_c + f(\hat{q_{12}}, \hat{q_{23}}, \hat{q_{13}}), \quad (11)$$

where $d_a$, $d_b$ and $d_c$ are the slopes.

Then, the interval with endpoints

$$\max_1 = f(q_{12} + \epsilon_{12}, q_{13} + \epsilon_{13}, q_{23} - \epsilon_{23}) \quad \text{and}$$
$$\min_1 = f(q_{12} - \epsilon_{12}, q_{13} - \epsilon_{13}, q_{23} + \epsilon_{23})$$

has confidence $c$ or higher. □

In our case, we are using the $f$ functions given in Lemma 1 to get the intervals for $\hat{p_1}$, $\hat{p_2}$ and $\hat{p_3}$. Since these functions are differentiable and we are only interested in values near $\hat{q_{12}}$, $\hat{q_{23}}$ and $\hat{q_{13}}$, the linearity assumption is reasonable.

**Summary: Evaluation with 3 Differences Scheme.** Consider three workers that perform $n$ tasks. First compute auxiliary estimates $\hat{q_{12}}$, $\hat{q_{23}}$ and $\hat{q_{13}}$ using the number of times each worker pair disagrees. For instance, $\hat{q_{12}} = b_{12}/n$ where $b_{12}$ is the number of times workers 1 and 2 disagree. Second, compute estimates for the three worker error rates $\hat{p_1}$, $\hat{p_2}$ and $\hat{p_3}$ using Lemma 1. Third, compute the half-sizes of the three confidence intervals $\epsilon_1$, $\epsilon_2$ and $\epsilon_3$ for the $p$ estimates using Lemma 3. □

### 3.1 Real Data Experiment

In this section we experimentally evaluate our confidence intervals, checking if they are valid, even if we do not know if the assumptions made hold. We use three real datasets to test our techniques : image comparison, predicate evaluation, and peer evaluation. In all these datasets, to estimate the accuracy of our confidence intervals, we use or estimate true answers for each of the tasks in the dataset. Note that the number of tasks is relatively small on purpose; one typically wants to evaluate workers after a small number of tasks, to save time and money.

**Data and Setting:** In the first scenario, which we call 'Image Comparison', denoted IC, a worker is asked to compare a pair of sports photos (each photo showing one athlete) and to indicate if the photos show the same person (Yes/No answer). The correct response for each task was known to us in advance. We evaluated the set of 48 tasks using 19 workers on Amazon's Mechanical Turk [1], accumulating a total of $48 \times 19$ responses. The data is available at [6].

The second dataset, which we call 'Schools of Thought', denoted SOT, is from [29]. A total of 402 workers were given 5 sets of 12 binary tasks each, making 60 tasks total per worker. In each task, workers were given an image, and asked to filter the image based on some question, such as 'Is it an image of the Sky?', or 'Is it an image of a building?', or 'Is the image beautiful?'. The correct response of each task is assumed to be the majority response for the task.

The third dataset is a MOOC (Massive Open Online Course) [5] grading dataset from the HCI Course at Stanford in Fall 2012, denoted MOOC. Students were asked to grade assignments of their peers, providing a rating from 0 to 5. Prior this peer grading process, each student was required to grade a fixed set of 'test' tasks. There were 388 such 'test' tasks that were evaluated by all workers (students). Since these 'test' tasks were also graded by staff members, the correct grade for each of these tasks is known to us. Since the grade provided is not binary, we interpret a grade from 0 to 2 as a *no* (or 'fail') and 3 to 5 as a *yes* (or 'pass').

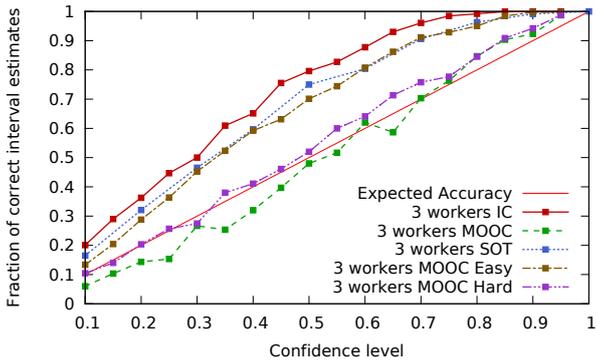

**Figure 1: Accuracy of $3$ worker differences method in estimating confidence**

In each of the datasets, we find that some tasks that had high agreement (100% of the workers returned the same answer), and some had low agreement (52% of the workers returned the same answer). Therefore, error probabilities of workers are positively correlated with each other (as opposed to zero correlation as required by our model), with all workers committing more errors on hard tasks. Moreover, we find that the assumption of false positive and false negative error rate both being equal is also not true in these datasets. We now want to see how good our 3-differences confidence interval estimates are for these data sets.

**Experiment:** We first pick a confidence level $c$. We then pick a random set of $m = 3$ workers, and run the general differences scheme on their responses (on tasks which all of them have attempted), to get a $c$-confidence interval for each of their error rates. We want to check if the worker's true error rate lies in the our $c$-confidence interval. Since we do not have access to the true worker error rate, we compute the fraction of errors made by each worker (using the correct responses, known to us), and use that as a proxy for the true error rate. We repeat this process for 100 random combinations of $m = 3$ workers, for every value of confidence level $c \in \{0.05, 0.1, 0.15, ..., 0.95\}$. For each confidence level $c$, we count the number of times the true error rate lay within our $c$-confidence interval, and divide by the number of confidence intervals we computed, and plot this value against $c$ in Figure 1. (The figure also contains two lines named '3 worker MOOC easy' and '3 worker MOOC hard'. They can be ignored for now.)

Note that the solid line $y = x$ in Figure 1 represents the ideal case. If an outcome is *above* the line, the predicted error rate was inside the computed interval *more than* $c$ percent of the time. Thus, these points are safe but conservative estimates. If an outcome is *below* the $y = x$ line, the confidence interval was too small. The closer an outcome is to the $y = x$ line, the less imprecise it is.

Our results show that in all cases the outcomes are close to the $y = x$ line. For the IC and the SOT data sets, our estimates are always above the $y = x$ line and hence always conservative. For the MOOC data set, the confidence interval is a bit imprecise but only for confidences less than 0.7. Since one would expect the desired confidence level to be above 0.7 to be useful, we believe this imprecision is not critical.

Of course, if we know more about the tasks and workers, and can model dependencies, task difficulties, and so on, one can do better.

To illustrate, we jump ahead and use one of the generalizations of Section 7. In the MOOC scenario, say we postulate that there are two types of tasks, hard and easy, and that workers perform differently on each type of task. Say we call tasks 'easy' if more than 90% of the workers that attempted them agree on the outcome, and 'hard' otherwise. (There are much more sophisticated ways to label tasks, but this is just an illustration.) We can then evaluate the error rates of workers for each type of task, using our same 3-differences approach. The results are shown in Figure 1. The new plots are labeled '3 worker MOOC easy' and '3 worker MOOC hard'. As the plot shows, the accuracy is now above the $y = x$ line, which suggests that our technique for handling difficulty works well in this case.

In summary, in scenarios where we do not have detailed information on correlations, biases, task difficulty, etc., our base-case 3-differences scheme seems to provide reasonable confidence intervals. If more information is available or learned, one can extend our method to take the information into account.

Although we have not shown the actual confidence intervals in Figure 1, it turns out some of them can be relatively large when one evaluates them over a small number of tasks. We have addressed this in Section 6.2.1

## 4. GENERAL DIFFERENCES SCHEME

Our solution for the differences scheme with 3 workers does not easily generalize to more than 3 workers. We could define $q_{ij}$ values for every possible pair of workers (analogous to Equations 1 through 3). From these values we can formulate equations for the worker error rates. If we have $j$ workers, we get $j(j-1)/2$ equations for $j$ desired rates $\hat{p_1}$ through $\hat{p_j}$. However, it is not clear how we may use all these equations to get the error rates.

Instead, for the $j$-worker case we use the following strategy. Say we are evaluating worker 1. Let us call the remaining $j-1$ workers the *peers*. From the peers we form two disjoint sets of workers, $S$ and $T$, and treat each of these sets as a "super-worker". That is, the result produced by the $S$ super-worker is the majority result of the $S$ workers, and similarly for the $T$ workers. We then apply the 3-worker solution to workers 1, $S$ and $T$ to get the desired error estimate $\hat{p_1}$.

The key question is what workers should go into sets $S$ and $T$. In the 3-worker solution, the accuracy of $\hat{p_1}$ will improve as the error rate of $S$ and $T$ decrease (Please refer to the technical report [16] for additional details on this). Thus, the question is how to obtain low-error rate $S$ and $T$ super-workers.

To illustrate the issues, say we have $j = 7$ workers and are evaluating worker 1. Assume the true error rates for the peers are $p_2 = p_3 = 0.1$ and $p_4 = p_5 = p_6 = p_7 = 0.4$. That is, peers 2 and 3 are good, and the rest are not very good. Say we form $S$ out of peers 2, 4 and 5. It is straightforward to compute the error rate of $S$ (the probability that the majority choice is incorrect) to be 0.208. In this case, we would have been better off simply having worker 2 alone in $S$! On the other hand, it is easy to construct cases where the super-worker has a lower rate than its individual members.

Hence, we need a strategy for selecting good $S$ and $T$ sets, i.e., ones that have the lowest (or close to lowest) possible error rates. Keep in mind that the selection procedure will not have access to the true error rates, as in our example.

We suggest three possible strategies for the $S, T$ selection. (Other variations are possible.)

- *Exhaustive.* We can try all possible disjoint $S, T$ possibilities. In practice, we expect $j$ to be relatively small, so we will not have an inordinate number of cases. (For example, for $j = 5$ there are 15 cases to consider.) For each $S, T$ choice, we estimate $\hat{p_1}$ and its accuracy using the 3-worker solution. At the end we choose the $\hat{p_1}$ value with best accuracy (smallest half-size of the confidence interval).

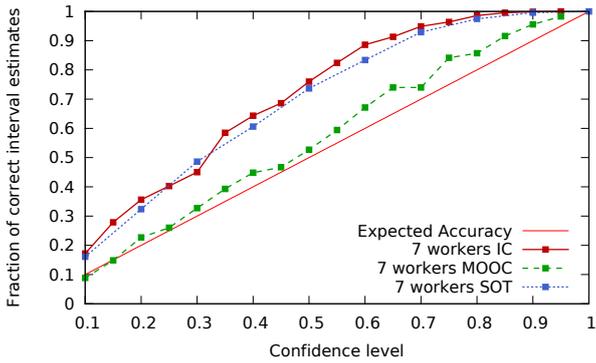

**Figure 2: Accuracy of multiple worker differences method in estimating confidence**

- *Pruning.* First, for each worker we compute a *preliminary* error rate by dividing its peers evenly into $S$ and $T$ sets. Then, to compute the final rate for a worker, we use the exhaustive method, but only considering peers that have a preliminary estimate below some threshold.
- *Greedy Search.* To evaluate worker $i$, pick a single random peer for $S$, and another one for $T$, and estimate error rates. Then consider adding each of the remaining peers, one at a time, to the set that has the largest estimated error. If adding the peer improves the accuracy for worker $i$, we add the peer to the set, otherwise we do not.

**Summary: Evaluation with General Differences Scheme.** Consider $j$ workers that perform $n$ tasks. To evaluate a worker $w$, divide the remaining workers into disjoint sets $S$ and $T$, and treat each set as a 'super-worker' whose response equals the majority response of workers from that set. Then proceed as in the 3 worker differences scheme, using worker $w$ and the two super workers. □

### 4.1 Real Data Experiment

We use the same three datasets we used to test the 3 worker differences scheme.

**Experiment:** Figure 2 shows the results for our evaluation of the general differences scheme. The data sets are as before, and the axis of the figure are the same as for Figure 1. In this case we use $m = 7$ workers instead of three, and we use a greedy heuristic to find super-workers. In this case we see that all outcomes are above the $y = x$ line, meaning that all our intervals are correct but conservative, even though some of our assumptions do not hold.

## 5. COMPARISON WITH EM AND SIMPLE MAJORITY

In this section, we compare our basic worker error estimates against the popular Expectation Maximization algorithm [8, 7], or EM, for short, and against a simple Majority Heuristic. The EM algorithm is used to provide estimates of hidden parameters, given values of observed parameters, such that the estimates are Maximum-Likelihood. In our setting, the hidden parameters are the worker error rates, which are unknown, while the observed parameters are the answers of workers in a given phase. Note, however, that the EM algorithm provides no confidence intervals for the hidden parameter estimates, unlike our method. So, for this comparison, we ignore our confidence intervals.

The EM algorithm uses two steps, computing the expectation (E), and maximization (M), until convergence. Specifically, the algorithm begins by initializing a random error rate for each worker and a random probability for Boolean answers for each task. Then, in alternate steps, it sets the maximum likelihood error rates for each worker based on the current answer probabilities, and sets the maximum likelihood answer probabilities given the worker responses and current error rates. The algorithm is guaranteed to converge to a local optimum of Maximum-Likelihood; however, in most practical settings, the algorithm does converge to the global optimum.

| Tasks | Workers | EM Error | Our Error | Simple Majority Error |
|---|---|---|---|---|
| 200 | 3 | 0.0552 | 0.0553 | 0.0638 |
| 300 | 3 | 0.0415 | 0.0345 | 0.0626 |
| 400 | 3 | 0.0355 | 0.0345 | 0.0626 |
| 500 | 3 | 0.0309 | 0.0299 | 0.0625 |
| 200 | 5 | 0.0330 | 0.0360 | 0.0344 |
| 300 | 5 | 0.0264 | 0.0282 | 0.0322 |
| 400 | 5 | 0.0235 | 0.0251 | 0.0307 |
| 500 | 5 | 0.0208 | 0.0224 | 0.0303 |

**Table 1: Comparison of our technique with EM**

The Simple Majority Heuristic [11], assumes that the majority response of workers is always correct, and hence evaluates worker's by counting the fraction of times they agreed/disagreed with the majority.

We tabulate the comparison between our algorithm (without confidence intervals), the EM algorithm, and the Simple Majority Heuristic in Table 1: For this comparison, we focused on a single phase, while varying the number of tasks (between 200-500), and while varying the number of workers (3 or 5). We generate task outcomes by assuming each worker has an error rate equal to 0.2 or 0.3 with probability $\frac{1}{2}$ each, independently of other workers. We then estimated the error rates $\hat{p}$ for each of the workers for a single phase using our technique, EM and Simple Majority, (ignoring confidence intervals), and recorded the error $|p - \hat{p}|$, i.e., how far away the error rate estimates are from the actual error rate. We recorded this value across 500 iterations, and then took the average.

As can be seen in the table, the error for EM and our technique is usually very close, while the error for simple majority is somewhat larger. For example, for 400 tasks and 3 workers, the average error for our technique is 0.034, that for EM is 0.035, while average error for the simple majority heuristic is 0.062. Thus, for all practical purposes, our technique and EM are equivalent to each other, and better than simple majority, in providing $\hat{p}$ estimates. Note that our technique, unlike the others, provides a wealth of more information in our confidence interval estimates, allowing us to better judge if workers are indeed of high quality.

## 6. USING OUR RESULTS

In this section we present case studies of how our confidence interval estimates may be used in various ways; while we have selected some especially important applications, we believe there may be many other ways this information may be used, some of which we haven't even envisioned yet.

### 6.1 Accuracy of Results

So far we have focused on estimating the error rates of workers. However, once we have worker error estimates, they can be used to improve the accuracy of the task results they produce.

For instance, say we have five workers, $W_1$ through $W_5$, that have executed a particular task. Say workers $W_1$, $W_2$ and $W_3$ all give the result Y, while workers $W_4$ and $W_5$ say N. Typically we would take a majority vote of the results, to mask out incorrect answers. In this case, our final task result would be Y, the majority vote. However, say that in the current phase we have evaluated the workers and we obtained that $\hat{p_1} = \hat{p_2} = \hat{p_3} = 0.4$ and that $\hat{p_4} = \hat{p_5} = 0.1$. Intuitively, it seems that we should weight the results of $W_4$ and $W_5$ more heavily since they are much more accurate, giving a final result of N.

We start by formally defining the variables used. Let random variable $X$ represent the correct result of a given task, either 1 if the result is Y or -1 if the result is N. We have $j$ workers and let random variable $X_i$ represent worker $i$ and $x_i$ (lowercase) be the specific result (1 or -1) given by worker $i$ ($1 \leq i \leq j$). Say we know exactly the worker error rates $p_i$. (Below we discuss what happens when we only have estimates $\hat{p}_i$.) Assume further that the task selectivity is $s$.

We wish to give a specific result $\hat{X}$ (either $\hat{X} = 1$ or $-1$) based on the results given by the workers. The *accuracy* of our result is the probability that $X = \hat{X}$ given the evidence (given the each $X_i = x_i$). For instance, if we say $\hat{X} = 1$ then the accuracy is the probability that $X = 1$ given the evidence. Our next lemma tells us how to compute the accuracy.

**Lemma 4.** To compute accuracy, we first compute two probabilities:

$$P_1 = \Pr[X = 1 \wedge X_1 = x_1 \wedge X_2 = x_2...]$$
$$= s \prod_{i=1}^{m}(p_i)^{\frac{1-x_i}{2}}(1-p_i)^{\frac{1+x_i}{2}} \text{ and}$$
$$P_{-1} = \Pr[X = -1 \wedge X_1 = x_1 \wedge X_2 = x_2...]$$
$$= (1-s) \prod_{i=1}^{m}(p_i)^{\frac{1+x_i}{2}}(1-p_i)^{\frac{1-x_i}{2}}.$$

Then, the accuracy of the task is:

- If $\hat{X} = 1$ then $\Pr[X = 1|X_1 = x_1 \wedge X_2 = x_2...] = \frac{P_1}{P_1 + P_{-1}}$, else
- If $\hat{X} = -1$ then $\Pr[X = -1|X_1 = x_1 \wedge X_2 = x_2...] = \frac{P_{-1}}{P_1 + P_{-1}}$. □

If we use a majority vote to compute the final task result $\hat{X}$, we can now say how accurate the answer is. However, our next lemma tells us how to select $\hat{X}$ by appropriately weighting the individual results, to maximize accuracy.

**Lemma 5.** We wish to give a specific result $\hat{X}$ (either $\hat{X} = 1$ or $-1$) based on the results given by the workers, such that we maximize accuracy, i.e., the probability that $X = \hat{X}$. To compute this maximum likelihood result we first compute

- $\alpha = log(\frac{s}{1-s})$,
- $\beta = \sum_{i=1}^{j} x_i log(\frac{1-p_i}{p_i})$.

Then, if $\alpha + \beta > 0$ the maximum likelihood result $\hat{X}$ is Y (1), else it is N (0).

Thus to provide a result with maximum accuracy, we need to give a weight $log(\frac{1-p_i}{p_i})$ to the answer of worker $i$, take the weighted sum, and add the initial bias factor $log(\frac{s}{1-s})$, and return the sign of the result as our guess $\hat{X}$. □

To illustrate the potential improvements that may be achieved by using Lemma 5, we consider a simple scenario with 9 workers. In this scenario we have two types of workers: good workers have a true error rate of 0.1, while bad workers have a rate of 0.3. Figure 3 shows the fraction of incorrect results (1 minus accuracy) for both a simple majority and a weighted majority (Lemma 5), as the number of bad workers varies. For instance, if we have 6 bad workers, the error rate for the combined result drops by about half when the optimal weighted majority is used (from about 0.02 to about 0.01). We see that the use of a weighted majority improves accuracy significantly, unless most workers are equally good, or unless every single worker is equally bad.

In practice we do not have the exact worker errors $p_i$ but only their estimates $\hat{p}_i$. To compute accuracy, we can use the estimates in place of the $p_i$ values in Lemma 4, to obtain *an estimate of the accuracy*. Thus, for each task we have $\hat{A}$ and not the true accuracy $A$. Is our estimate for $\hat{A}$ any good? If we tell whoever is consuming the results of our tasks that "we think that" the answer we provide is correct 95% of the time, how will they know if they can trust that 95% accuracy estimate?

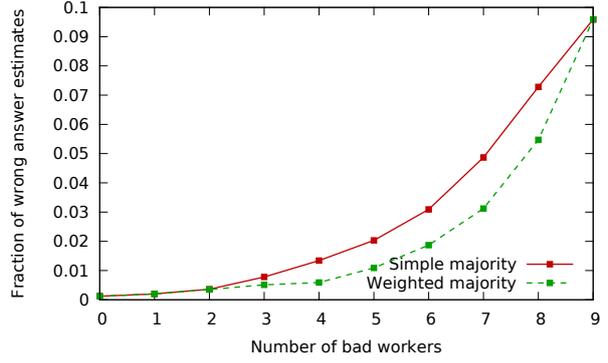

**Figure 3: Fraction of incorrect results using a simple vs. a weighted majority**

Fortunately, not only do we have error estimates for the workers, but we also have confidence intervals for those estimates! Thus we can perform a "worst-case" analysis: For each $\hat{p}_i$ we know with $c$ confidence level that at most the true $p_i$ is $\hat{p}_i + \epsilon_i$. We now estimate $\hat{A}'$, which is the accuracy we obtain when we use $\hat{p}_i + \epsilon_i$ instead of $p_i$ in Lemma 4. Since $\hat{A}'$ is a differentiable function, and since larger error rates make accuracy worse (smaller), we can say that with $c$ confidence, the true accuracy $A$ will be *larger* than $\hat{A}'$. Note that once again, we are using the linearity principle of Lemma 3, i.e., that the majority function for $\hat{A}'$ is locally linear around the estimates of the values of $\hat{p}_i$.

Returning to our simple example, say our (worst case) accuracy bound is 95% with 90% confidence. Now we can tell our customer that over many tasks we will give an incorrect answer in no more than 14.5% of the cases, since we do not lie within the desired $\hat{A}'$ accuracy with probability 0.1, and if we do lie within the desired accuracy bound (with probability 0.9), then with probability 0.05 we end up making an error, giving us overall: $(1 - 0.9) + 0.9 \times (1 - 0.95) = 0.145$.

## 6.2 Price of Accuracy

In the previous subsection, we provided a technique for providing a confidence value along with an accuracy bound for the answer for any given task: we used the larger extreme of the worker error probabilities $\hat{p}_i + \epsilon_i$ to provide an accuracy bound $\hat{A}'$. Furthermore, we know that the worker error probabilities $p_i$ are all smaller than $\hat{p}_i + \epsilon_i$ with confidence $c$. Thus, overall, we get an accuracy bound of $\hat{A}'$ with confidence value $c$.

In this section, we first set a lower bound on the desired accuracy, and we study the impact on the confidence value of varying the number of tasks or number of workers in a single phase. Note that increasing the number of tasks will increase the confidence value, as will increasing the number of workers.

In Figure 4(a), we plot the number of workers required to get answer accuracy of at least 90% at different confidence levels on fixing the number of tasks to 500. As predicted, as the confidence levels increase, the number of workers required increases. In Figure 4(b), we plot the number of tasks required to get answer accuracy of 90% at different confidence levels on fixing the number of workers to 3. Once again, as predicted, as the confidence levels increase, the number of tasks required increases. In a sense,

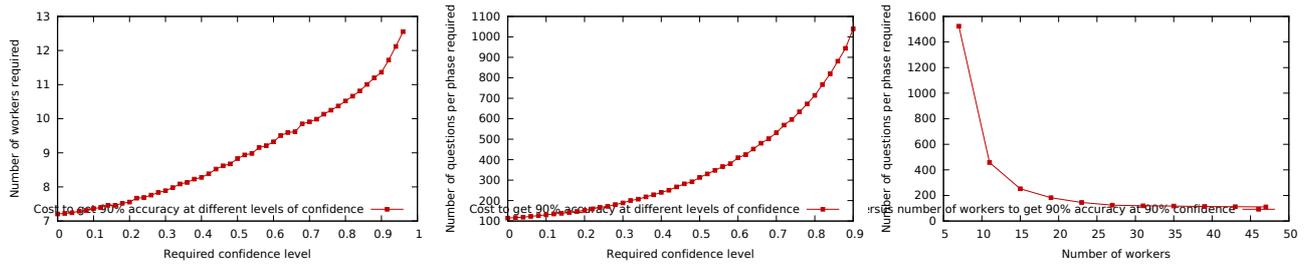

**Figure 4: (a)** Effect of number of workers on confidence level of answers **(b)** Effect of number of tasks on confidence level of answers **(c)** Tradeoff between number of tasks and number of workers on fixing confidence level of answers

these graphs illustrate the "cost of confidence/accuracy". That is, the graphs show us how much we need to "pay", in terms of more workers or more tasks to achieve our goal.

Lastly, in Figure 4(c), we plot the dependence of the number of workers on the number of tasks when we fix the confidence level to 90%, and the answer accuracy of at least 90%. In other words, we fix the number of workers, and increase the number of tasks until we get the desired confidence and accuracy levels. We may use this plot to find the optimal tradeoff between the number of tasks ($n_d$) and number of workers ($w$). Let us say we wanted to ensure the confidence level and answer accuracy of a single task to be both greater than 90%. Then, to ensure this accuracy level, we can use various combinations of $n_d$ and $w$ with the total cost for a phase being $n_d \times w$. For $w = 11$ in the figure, we find that $n_d \approx 450$, while for $w = 15$, $n_d \approx 250$. We find that the optimal combination of $n_d$ and $w$ occurs around $w = 25$, where $n_d \approx 130$, for a total number of task instances of $n_d \times w = 3250$.

### 6.2.1 Why Confidence Intervals?

Figures 4(a), 4(b) seem to indicate that getting small confidence intervals, for high levels of confidence, requires a large number of tasks and workers. For smaller amounts of data, the interval sizes can be relatively large. This is partly because schemes based on worker differences (as opposed to gold standard tasks) usually require more data to get the same-sized confidence intervals. So the reader may wonder, how useful are these intervals if they can be large for a modest number of tasks or workers?

First, simply ignoring the intervals (as is commonly done) is not a good idea even if the intervals are large. (Recall the proverbial ostrich hiding its head underground.) Large intervals give us a warning that our estimates may be incorrect, and it is best to take that into account. For instance, in Section 6.3 we will see that it is good to take into account uncertainty when evicting workers. (When we evict workers with uncertain error estimates, it is better to be conservative than overly aggressive.) We can also heed the warning by evaluating workers over more tasks, changing our evaluation scheme (to say using a gold standard), or by having more workers repeat each task. Whether we can afford more tasks or workers is an orthogonal issue that can only be discussed rationally with knowledge of confidence intervals.

Second, note that computing confidence intervals is "free." That is, knowing the confidence intervals does not require additional work beyond what is already being done to evaluate workers and to perform tasks.

Third, also note that even wide confidence intervals can be useful if we are trying to distinguish between classes of workers that have quite different rates. For instance, if we are trying to detect spammers (rates close to 0.5) from good workers (say with rates below 0.05), then coarse intervals can easily differentiate the workers.

## 6.3 Multiple Phases

As discussed in the introduction, we may also use our error estimates to periodically evict poorly-performing workers. Prior work [24, 11] has used heuristics to eliminate poorly performing workers, but here, armed with our toolbox of evaluation techniques, we may use precise estimates of worker errors to truly judge if a worker is good or bad.

Specifically, at the end of every phase, we choose to evict some workers based on how well they performed in that phase, and replace them with other workers. A straightforward technique is to simply use our estimate of the error rate of the worker ($\hat{p}$), and reject workers whose error rates fall above an appropriately set threshold $t$. We call this technique the *normal eviction* technique.

Instead, we may use our confidence interval estimates to only evict workers who are truly poor. That is, we may use $\hat{p}_i - \epsilon$ as an estimate of the lowerbound of the error rate of a worker. If the lowerbound is still higher than an appropriately set threshold $t$ with confidence $c$, we can be confident that the worker is indeed poor and can then evict him/her. We call this technique the *conservative eviction* technique. Note that $\epsilon$ varies for each worker, i.e., for some workers we can be more precise about their error, and for other less so. Thus, conservative eviction sets an error threshold that varies by worker, taking into account our precision.

Since some workers may be retained across phases, we estimate error rates across multiple phases. That is, we compute the average rate over all phases a worker has been involved in, and then apply the threshold. (There are other options for averaging but are not considered here.)

To illustrate our techniques, consider a scenario where a newly hired worker will have error rate 0.3 with probability 0.3, 0.2 with probability 0.4, and 0.1 with probability 0.3. Note that identifying and evicting poor workers in such a scenario is especially important since the error rates of the good and bad workers are very different.

Our goal is to compare normal and conservative eviction on two metrics: (a) how poor our workers are at the end of each phase, and (b) how many good workers do we mistakenly evict. Our first metric prefers techniques that select for good workers, while the second metric prefers techniques that are not overly aggressive. Note that mistakenly evicting good workers can result in a poor reputation for the requester in the crowdsourcing marketplace, and can have detrimental effect on quality in the long run (since good workers will refuse to work with the requester). Note also that the two aspects are interrelated, e.g., aggressive eviction may improve quality, but may evict many good workers.

To evaluate these aspects, we use a simple accuracy cost function $C = c_1 + \alpha c_2$: The value $c_1$ evaluates how bad the workers are at the end—we assign a cost of 1 for ending up with a worker with error rate 0.2, and a cost of 3 for a worker with error rate 0.3 (and a cost of 0 for a worker with error rate 0.1). The value $c_2$ evaluates how many good workers do we mistakenly evict—we assign a cost of 5 for each worker with error rate 0.1 evicted at the end of a phase, and a cost of 0 if we evict workers with error rates 0.2 or 0.3. The parameter $\alpha > 0$ is a multiplier that allows us to weight $c_1$ and $c_2$ relative to each other. Note that these costs are merely illustrative: in a real application, we may use different functions for $c_1$ and $c_2$, and different $\alpha$s.

In our experiment, we evaluate $C$ for both techniques at the end of every phase, and then take the average across phases. We fix the number of phases $k$ at 30, with the number of tasks in each phase $n$ being 25. (Similar results were obtained for $k = 50, 75, 100$, with $nk$ fixed.)

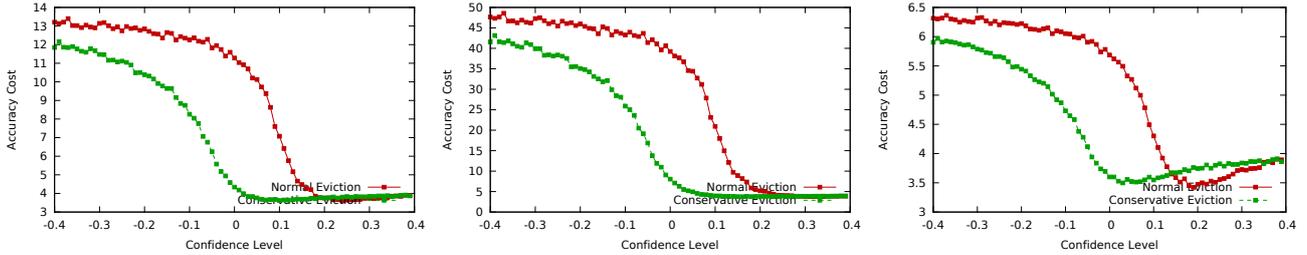

Figure 5: (a) Accuracy Cost vs. Threshold of Conservative and Normal Eviction for $\alpha = 1$ (b) Accuracy Cost vs. Threshold of Conservative and Normal Eviction for $\alpha = 5$ (c) Accuracy Cost vs. Threshold of Conservative and Normal Eviction for $\alpha = \frac{1}{5}$

In Figure 5(a), we depict the average accuracy cost $C$ (where $\alpha = 1$) in log scale as a function of threshold $t$, for both eviction techniques, and for 35% confidence intervals (Similar behavior is observed for 50% and 75% intervals.) Note that we consider negative threshold values, even though error probabilities can only be positive. Our techniques can yield negative estimates, e.g., $\hat{p}_i - \epsilon$ can be negative. We find that it is more effective not to cut off these values at 0, since the magnitude still conveys some useful information (e.g., we would rather evict a worker with lowerbound -0.2 than one with a bound of -0.1).

As can be seen in the figure, the average cost of conservative eviction is typically lower than normal eviction across all thresholds (except for a small region between 0.2 and 0.3 where the difference is not much). Moreover, we find that on both sides of 0, i.e., from $0 \to 0.4$ and from $0 \to -0.4$, the accuracy cost grows much slower for the conservative technique than the normal technique. The slower growth implies that conservative is less sensitive to the threshold choice than normal. Thus, even though at their optimal thresholds both techniques perform roughly equally, if we are unable to set the threshold precisely, conservative will do better.

To study the impact of $\alpha$ on performance, we repeated the experiment with $\alpha = 5$ and $\alpha = \frac{1}{5}$, depicted in Figures 5(b) and 5(c) respectively. We note similar behavior for both plots. For large $\alpha$, the curve is almost monotonically decreasing—the trough is almost not visible, unlike $\alpha = 1$ or $\frac{1}{5}$—this behavior is not surprising given that if we put too large a penalty on evicting good workers, then the best strategy is to simply not evict any workers. Also, for both figures, we find that conservative eviction has lower cost than normal eviction for almost every threshold (except for $\alpha = \frac{1}{5}$ between 0.17 and 0.35, where the difference is not very large). Just as in Figure 5(a), the accuracy cost growth of conservative eviction is slower than normal eviction on both sides of the origin—thus even if we happen to not choose the optimal threshold, conservative eviction will ensure that we end up not paying as much of a price in terms of accuracy cost.)

To summarize, our confidence interval estimates provide a useful basis to evaluate and maintain worker quality across phases, while simultaneously ensuring good quality and few evictions of good workers.

## 7. EXTENDING OUR TECHNIQUES

Our base methods assume a relatively simple model, where task outputs are binary, and worker error rates are not dependent on task difficulty or type. We also assume that task selectivity is unknown. If we have more information, we can obtain more refined confidence intervals by extending our base mechanism. Here we briefly discuss four such extensions. Other more sophisticated extensions are possible, but not covered here due to space limitations.

**Selectivity:** In some cases, the task *selectivity* $s$ may be known beforehand. That is, we may know in advance that with probability $s$ the result of given task is Y, and with probability $(1-s)$ it is N. In this case, we can construct a ''worker' who always responds with 'yes' if $s > 0.5$ and no if $s < 0.5$. This worker will have an error rate of $min(s, 1-s)$. By considering this worker along with two or more additional workers, we can use the 3 worker or multiple worker differences scheme, to get error rates for the workers. The differences method gives more precise results when more data is available [16], and so using selectivity can improve our worker evaluation by providing us with the responses of an extra 'worker'.

**Non-binary tasks:** Some tasks may have non-binary outcomes. For example, if the task is to identify the background color in a photograph, we may have outcomes red, blue, green, and yellow. However, non-binary tasks can be reduced to multiple binary tasks, which can then be analyzed by our methods. To illustrate, say the output of a task is one of $k$ categories. If $k$ is a power of two, then we can express the category as a $log(k)$ length binary string, and convert the task to $log(k)$ binary tasks of the form : Is the $i^{th}$ bit of the category 1? For example, for the four colors red, blue, green and yellow, we can map (for our analysis) each worker answer to two answers for the questions 'Is the background either red or green?' and 'Is the background either yellow or green?'. If $k$ is not a power of two, we can add more categories until it becomes a power of two. In our example, if the picture backgrounds were all red, green or blue, we can still add another category yellow. The workers will never reply yellow, but our methods can still be applied to the results.

**Multiple Task Types:** Typical crowdsourcing marketplaces have tasks of various types, for example, translation (language centric), identifying people in images (knowledge centric), or debugging code (programming centric). A worker may have a different aptitude level for each task type, and hence a different error rate, violating our assumption of a single error rate. To account for multiple task types, we can apply our method to tasks of only one type at a time, and find a worker's error rate separately for each task type.

**Varying Task Difficulty:** Tasks in a set may have varying difficulties, and the error rate of a worker may be higher for more difficult tasks. If we can identify difficulty level of tasks beforehand, then we can treat tasks of different difficulty level the way we treated tasks of different type. If we don't know difficulty levels beforehand, we can try to deduce them by looking at the strength of the majority of worker responses. For instance, a majority close to 100% would indicate an easy task, while majority close to 50% would indicate a hard one. We illustrated in Figure 1 the use of this technique, and the improvements it yields.

## 8. RELATED WORK

The prior work related to ours can be placed in four categories; we describe each of them in turn:

**Crowd Algorithms:** There has been a lot of recent work on finding crowdsourcing analogs of standard data processing algorithms, such as filtering [21], sorting and joins [14, 19], deduplication and clustering [3, 23, 31] and categorization [22, 26]. Most of these algorithms assume a simple model of human errors (i.e., that all human beings are alike.) All these algorithms would benefit from a

"prefiltering" or "evaluation" phase where humans with low accuracy not employed for tasks.

**Statistics and EM:** Expectation Maximization, or EM [7, 8] has been studied and used in the statistics and machine learning community for several decades now, with many textbooks and surveys on the topic [15, 20, 30]. Expectation Maximization provides maximum likelihood estimates for hidden model parameters based on a sequence of steps that converge to a locally optimal solution. We compare our approach against Expectation Maximization in Section 5. However, EM, unlike our technique, does not provide confidence intervals for error rate estimates.

There is a variant of the EM algorithm called multiple imputation [27], which can be used to derive a distribution of hidden model parameters. In our case, we derive confidence intervals for hidden model parameters. Unlike our method which provide correct guarantees, multiple imputation is heuristic in nature.

We use many fundamental concepts from statistics and probability [32, 4], including the Wilson score interval.

**Worker Error Estimation:** The work most closely related to ours is that of simultaneous estimation of answers to tasks and errors of workers (typically using the EM algorithm). There have been a number of papers studying increasingly expressive models for this problem, including difficulty of tasks and worker expertise [9, 13], adversarial behavior [25], and online evaluation of workers [33, 24, 18]. While our worker error model and task model are simpler, we provide confidence interval guarantees along with error rates, allowing users of our technique to have more fine-grained information to evaluate workers.

There has also been some work on selecting which items to get evaluated by which workers in order to reduce overall error rate [17, 28, 11]. While [11] uses heuristic confidence intervals in their algorithm, but does not provide confidence guarantees, the other papers do not provide confidence intervals of any kind.

**Applications:** There are a number of applications that crowdsourcing has been successfully used for, including sentiment analysis [12], identifying spam [10], determining search relevance [2], translation [35]. All these applications would benefit from a prefiltering phase where the workers are evaluated and the poor workers (judged based on average behavior as well as confidence intervals) can be barred from further work.

## 9. CONCLUSION

In this paper, we presented a technique for determining confidence intervals in addition to worker error rates. Confidence intervals provide the equivalent of "guarantees" in the setting where we need to determine worker errors, and thus are useful when we need to precisely determine the range in which worker errors may lie. We showed that our confidence intervals are relatively accurate even in scenarios where our assumptions do not hold, and we discussed extensions for scenarios where the assumptions are markedly different.